\documentclass[british,conference]{IEEEtran}
\usepackage[latin9]{inputenc}
\usepackage{array}
\usepackage{wrapfig}
\usepackage{url}
\usepackage{multirow}
\usepackage{amsmath}
\usepackage{amssymb}
\usepackage{graphicx}

\makeatletter

\providecommand{\tabularnewline}{\\}

 \usepackage[ruled,vlined]{algorithm2e}
 \usepackage{tikz}

\LinesNumbered
\let\oldnl\nl
\newcommand{\nonl}{\renewcommand{\nl}{\let\nl\oldnl}}

\@ifundefined{showcaptionsetup}{}{%
 \PassOptionsToPackage{caption=false}{subfig}}
\usepackage{subfig}
\makeatother

\usepackage{babel}
\begin{document}

\title{Smart Sampling for Lightweight Verification \\
of Markov Decision Processes}

\author{Pedro D'Argenio\IEEEauthorrefmark{1}, Axel Legay\IEEEauthorrefmark{2},
Sean Sedwards\IEEEauthorrefmark{2} and Louis-Marie Traonouez\IEEEauthorrefmark{2}\\
\IEEEauthorrefmark{1}Universidad Nacional de C\'ordoba, Argentina,
and \IEEEauthorrefmark{2}Inria Rennes -- Bretagne Atlantique, France}


\maketitle
\begin{abstract}
Markov decision processes (MDP) are useful to model optimisation problems
in concurrent systems. To verify MDPs with efficient Monte Carlo techniques
requires that their nondeterminism be resolved by a scheduler. Recent
work has introduced the elements of lightweight techniques to sample
directly from scheduler space, but finding optimal schedulers by simple
sampling may be inefficient. Here we describe ``smart'' sampling
algorithms that can make substantial improvements in performance.
\end{abstract}

\section{Introduction}

Markov decision processes describe systems that interleave nondeterministic
\emph{actions} and probabilistic transitions. This model has proved
useful in many real optimisation problems \cite{White1985,White1988,White1993}
and may be used to represent concurrent probabilistic programs (see,
e.g., \cite{BiancoDeAlfaro1995,BaierKatoen2008}). Such models comprise
probabilistic subsystems whose transitions depend on the states of
the other subsystems, while the order in which concurrently enabled
transitions execute is nondeterministic. This order may radically
affect the behaviour of a system and it is thus useful to calculate
the upper and lower bounds of quantitative aspects of performance.

As an example, consider the network of computational nodes depicted
in Fig. \ref{fig:network} (relating to the case study in Section
\ref{sec:virus}). Given that one of the nodes is infected by a virus,
we would like to calculate the probability that a target node becomes
infected. If we know the probability that the virus will pass from
one node to the next, we could model the system as a discrete time
Markov chain and analyse it to find the probability that any particular
node will become infected. Such a model ignores the possibility that
the virus might actually choose which node to infect, e.g., to maximise
its probability of passing through the barrier layer. Under such circumstances
some nodes might be infected with near certainty or with only very
low probability, but this would not be adequately captured by the
Markov chain. By modelling the virus's choice of node as a nondeterministic
transition in an MDP, the maximum and minimum probabilities of infection
can be considered.

\begin{figure}
\begin{minipage}[t]{0.5\columnwidth}%
\begin{tikzpicture}[every node/.style={draw,circle}]
\draw
(0,0)node(00){}
(1,0)node(10){}
(2,0)node(20){};
\draw
(0,1)node[fill=gray](01){}
(1,1)node[fill=gray](11){}
(2,1)node[fill=gray](21){};
\draw
(0,2)node(02){}
(1,2)node(12){}
(2,2)node(22){};
\draw(2.8,0)node[draw=none]{\smaller infected};
\draw(2.8,1)node[draw=none, align=left]{\smaller barrier\\[-2pt]\smaller layer};
\draw
(2,0)node[inner sep=1pt,fill=black]{}
(-0.6,2)node[draw=none]{\smaller target};
\draw(00)edge(01)edge(10)(01)edge(02)edge(11);
\draw(22)edge(12)edge(21)(10)edge(11)edge(20);
\draw(12)edge(02)edge(11)(21)edge(20)edge(11);
\end{tikzpicture}\caption{Model of network virus infection.\label{fig:network}}
\end{minipage}\qquad{}%
\begin{minipage}[t]{0.42\columnwidth}%
\centering
\begin{tikzpicture}[scale=0.9,every node/.style={draw,circle,inner sep=1pt,scale=0.9}]
\draw(0,0)node(0){$s_0$}(-1,-1)node[fill=black](1){}
(0,-1)node[fill=black](2){}(-2,-2)node(3){$s_1$}(-1,-2)node(4){$s_2$}
(0,-2)node(5){$s_3$}(1,-2)node(6){$s_4$};
\draw[->,every node/.style={circle,inner sep=1pt,scale=0.9}]
(0)edge[<-](0,0.75)
(0)edge[-]node[above left]{$a_1$}(1)
(0)edge[-]node[left]{$a_2$}(2)
(1)edge node[above left]{$p_1$}(3)(1)edge node[below left]{$p_2\,$}(4)
(2)edgenode[below left]{$p_3\,$}(5)(2)edgenode[right]{$~p_4$}(6);
\draw[every edge/.style={dashed,draw}]
(3)edge(-2.25,-2.5)(3)edge(-1.75,-2.5)(4)edge(-1.25,-2.5)(4)edge(-0.75,-2.5)
(5)edge(-.25,-2.5)(5)edge(0.25,-2.5)(6)edge(0.75,-2.5)(6)edge(1.25,-2.5)(1)edge(-0.5,-1.5)(0)edge(1,-1);
\end{tikzpicture}\caption{Fragment of a Markov decision process.\label{fig:MDP}}
\end{minipage}
\end{figure}

Fig. \ref{fig:MDP} shows a typical fragment of an MDP. Its execution
semantics are as follows. In a given state ($s_{0}$), an action ($a_{1},a_{2},\dots$)
is chosen nondeterministically to select a distribution of probabilistic
transitions ($p_{1},p_{2},\dots$ or $p_{3},p_{4}$, etc.). A probabilistic
choice is then made to select the next state ($s_{1},s_{2},s_{3},s_{4},\dots$).
In this work we use the term \emph{scheduler} to refer to a particular
way the nondeterminism in an MDP is resolved.

Classic analysis of MDPs is concerned with finding the expected maximum
or minimum reward for an execution of the system, given individual
rewards assigned to each of the actions \cite{Bellman1957,Puterman1994}.
Rewards may also be assigned to states or transitions between states
\cite{KwiatkowskaNormanParker2007}. In this work we focus on MDPs
in the context of \emph{model checking} concurrent probabilistic systems,
to find schedulers that maximise or minimise the probability of a
property. Model checking is an automatic technique to verify that
a system satisfies a property specified in temporal logic \cite{ClarkeEmersonAllenSifakis2009}.
\emph{Probabilistic} model checking quantifies the probability that
a probabilistic system will satisfy a property \cite{HanssonJonsson1994}.
Numerical model checking algorithms to solve purely probabilistic
systems are costly in time and space. Finding extremal probabilities
in MDPs is generally more so, but is nevertheless a polynomial function
of the explicit description of the MDP \cite{BiancoDeAlfaro1995}.

\emph{Statistical} model checking (SMC) describes a collection of
Monte Carlo sampling techniques that make probabilistic model checking
more tractable by returning approximative results with statistical
confidence \cite{YounesSimmons2002}. Recent approaches to apply SMC
to MDPs \cite{Bogdoll2011,LassaignePeyronnet2012,Henriques-et-al2012,HartmannsTimmer2013,LegaySedwardsTraonouez2014}
are memory-intensive and do not fully address the nondeterministic
model checking problem. Classic sampling approaches for MDPs, such
as the Kearns algorithm \cite{KearnsMansourNg2002}, address a related
but different problem.

This work extends \cite{LegaySedwardsTraonouez2014}. In \cite{LegaySedwardsTraonouez2014}
the authors provide sampling techniques that can form the basis of
memory-efficient (``lightweight'') verification of MDPs. The principal
contributions of \cite{LegaySedwardsTraonouez2014} are (\emph{i})
specifying the infinite behaviour of history-dependent or memoryless
schedulers using $\mathcal{O}(1)$ memory, (\emph{ii}) sampling directly
and uniformly from scheduler space, and (\emph{iii}) quantifying the
statistical confidence of multiple estimates or multiple hypothesis
tests. As in the case of standard SMC, sampling makes the verification
problem independent of the size of the space of samples, with a convergence
to the correct result almost surely guaranteed with an infinite number
of samples. The use of lightweight techniques opens up the possibility
to efficiently distribute the problem on high performance massively
parallel architectures, such as general purpose computing on graphics
processing units (GPGPU).

Sampling schedulers makes a significant advance over mere enumeration.
For example, suppose half of all schedulers for a given MDP and property
are `near optimal', i.e., have a probability of satisfying the property
that is deemed adequately close to the true optimum. If all such near
optimal schedulers lie in the second half of the enumeration, it will
be necessary to enumerate half of all schedulers before finding one
that is near optimal. In contrast, one would expect to see a near
optimal scheduler after just two random selections, i.e., the expectation
with two samples is one. This phenomenon is not limited to the case
when schedulers are pathologically distributed with respect to the
enumeration. Since the total number of schedulers increases exponentially
with path length (the number of history-dependent schedulers increases
doubly exponentially with states and path length), the total number
of schedulers is usually very large. Hence, even when near optimal
schedulers are more uniformly distributed with respect to the enumeration,
it is typically not tractable to use enumeration to find one. Note
that sampling also works with non-denumerable spaces. The cost of
finding a near optimal scheduler with sampling is simply proportional
to the relative mass of near optimal schedulers in scheduler space.
Our experiments with standard case studies suggest that this cost
is typically reasonable.

It was demonstrated in \cite{LegaySedwardsTraonouez2014} that simple
undirected sampling may be adequate for some case studies. In this
work we present ``smart sampling'' algorithms that make significantly
better use of a simulation budget. For a given number of candidate
schedulers, smart sampling can reduce the simulation cost of extremal
probability estimation by more than $N/\lceil2+\log_{2}N\rceil$,
where $N$ is the minimum number of simulations necessary to achieve
the required statistical confidence, as given by (\ref{eq:NChernoff}).
The basic notions of smart sampling were hinted at in \cite{LegaySedwardsTraonouez2014}.
Simply put, a small part of the budget is used to perform an initial
assessment of the problem and to generate an optimal candidate set
of schedulers. The remaining budget is used to test and refine the
candidate set: sub-optimal schedulers are removed and their budget
is re-allocated to good ones. Here we give a full exposition of smart
sampling and explain its limitations. We have implemented the algorithms
in our statistical model checking platform, \textsc{Plasma}%
\footnote{project.inria.fr/plasma-lab/%
}, and demonstrate their successful application to a number of case
studies from the literature. We include some examples that are intractable
to numerical techniques and compare the performance of our techniques
with an alternative sampling approach \cite{Henriques-et-al2012}.
We also give an example where smart sampling is less effective, but
show that the results may nevertheless be useful in bounding the range
of extremal probabilities.

\subsection*{Structure of the Paper}

In Section \ref{sec:prelims} we introduce some basic concepts and
notation necessary for the sequel. In Section \ref{sec:related} we
briefly survey closely related work. In Sections \ref{sec:seeds}
we recall the basis of our lightweight verification techniques. In
Section \ref{sec:smart} we describe the notion of smart sampling
and present our smart estimation and smart hypothesis testing algorithms.
In Section \ref{sec:experiments} we give the results of experiments
with a number of case studies from the literature. In Section \ref{sec:counterexamples}
we discuss the limitations of smart sampling and in Section \ref{sec:prospects}
we summarise the challenges and prospects for our approach.

\section{Preliminaries\label{sec:prelims}}

Given an MDP with set of actions $A$, having a set of states $S$
that induces a set of sequences of states $\Omega=S^{+}$, a history-dependent
(general) scheduler is a function $\mathfrak{S}:\Omega\rightarrow A$.
A memoryless scheduler is a function $\mathfrak{M}:S\rightarrow A$.
Intuitively, at each state in the course of an execution, a general
scheduler ($\mathfrak{S}$) chooses an action based on the sequence
of previous states and a memoryless scheduler ($\mathfrak{M}$) chooses
an action based only on the current state. History-dependent schedulers
therefore include memoryless schedulers.

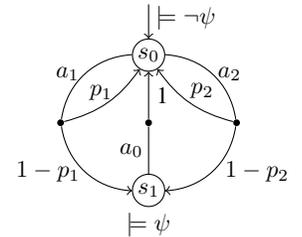
\begin{wrapfigure}{o}{0.47\columnwidth}%
\centering
\begin{tikzpicture}[scale=0.9,every node/.style={circle,inner sep=1pt,scale=0.9}]
\draw
(0,0)node[draw](0){$s_0$}
(0,-2)node[draw](3){$s_1$}
(0)node[above right=0.5em]{$\models\neg\psi$}
(3)node[rectangle,below=0.8em]{$\models\psi$};
\draw
(0)edge[<-](0,0.75)
(-1.3,-1)node[fill=black](2){}
(1.3,-1)node[fill=black](4){}
(0,-1)node(1)[fill=black]{}(3);
\draw[every edge/.style={->,bend right=39,draw}]
(0)edge[-]node[left]{$a_1$}(2)
(2)edgenode[left]{$1-p_1$}(3)
(4)edge[bend left=39]node[right]{$~1-p_2$}(3)
(0)edge[-,bend left=39]node[right]{$a_2$}(4)
(2)edge[bend right=20]node[above left]{$p_1\!\!$}(0)
(4)edge[bend left=20]node[above right]{$p_2$}(0);
\draw[<-]
(1)edge[-]node[left]{$a_0$}(3)
(0)edgenode[right]{1}(1);
\end{tikzpicture}\caption{MDP with different optima for general and memoryless schedulers.\label{fig:historydependent}}
\end{wrapfigure}%

Fig. \ref{fig:historydependent} illustrates a simple MDP for which
memoryless and history-dependent schedulers give different optima
for the bounded temporal logic property $\mathbf{X}(\psi\wedge\mathbf{XG}^{t}\neg\psi)$
when $p_{1}\neq p_{2}$ and $t>0$. The property makes use of the
temporal operators \emph{next} ($\mathbf{X}$) and \emph{globally}
($\mathbf{G}$). Intuitively, the property states that on the next
step $\psi$ will be true and, on the step after that, $\neg\psi$
will be remain true for $t+1$ time steps. The property is satisfied
by the sequence of states $s_{0}s_{1}s_{0}s_{0}\cdots$. If $p_{1}>p_{2}$,
the maximum probability for $s_{0}s_{1}$ is achieved with action
$a_{2}$, while the maximum probability for $s_{0}s_{0}$ is achieved
with action $a_{1}$. Given that both transitions start in the same
state, a memoryless scheduler will not achieve the maximum probability
achievable with a history-dependent scheduler.

\subsection*{Statistical Model Checking with \textsc{Plasma\label{sec:smc}}}

The algorithms we present here are implemented in our SMC platform
\textsc{Plasma} (Platform for Learning and Advanced Statistical Model
checking Algorithms \cite{Boyer-et-al2013}). \textsc{Plasma} is modular,
allowing new modelling languages, logics and algorithms to be plugged-in
and take advantage of its graphical user interface, its integrated
development environment and its ability to correctly divide simulations
on parallel computational architectures. We introduce here the basic
features of SMC with \textsc{Plasma} that are relevant to what follows.

SMC algorithms work by constructing an automaton to accept only traces
that satisfy a specified property. The state space of the system is
not constructed explicitly -- states are generated on the fly during
simulation -- hence SMC is efficient for large, possibly infinite
state, systems. Moreover, since the simulations are required to be
statistically independent, SMC may be efficiently divided on parallel
computing architectures. The automaton may then be used to estimate
the probability of the property or to decide an hypothesis about the
probability. Typically, the probability $p$ of property $\varphi$
is estimated by the proportion of traces that individually satisfy
it, i.e., $p\approx\frac{1}{N}\sum_{i=1}^{N}\mathbf{1}(\omega_{i}\models\varphi)$,
where $\omega_{1},\dots,\omega_{N}$ are $N$ independently generated
simulation traces and $\mathbf{1}(\cdot)$ is an indicator function
that corresponds to the output of the automaton: it returns $1$ if
the trace is accepted and $0$ if it is not. As a statistical process,
the results of SMC are given with probabilistic confidence.

In the case of estimation, \textsc{Plasma} calculates a priori the
required number of simulations according to a Chernoff bound \cite{Okamoto1958}
that allows the user to specify an error $\varepsilon$ and a probability
$\delta$ that the estimate $\hat{p}$ will not lie outside the true
value $\pm\varepsilon$. Given that a system has true probability
$p$ of satisfying a property, the Chernoff bound ensures $\mathrm{P}(\mid\hat{p}-p\mid\geq\varepsilon)\leq\delta$.
Parameter $\delta$ is related to the number of simulations $N$ by
$\delta=2e^{-2N\varepsilon^{2}}$ \cite{Okamoto1958}, giving
\begin{equation}
N=\left\lceil (\ln2-\ln\delta)/(2\varepsilon^{2})\right\rceil .\label{eq:NChernoff}
\end{equation}

In the case of hypothesis testing, \textsc{Plasma} adopts the sequential
probability ratio test (SPRT) of Wald \cite{Wald1945} to test hypotheses
of the form $\mathrm{P}(\omega\models\varphi)\bowtie\theta$, where
$\bowtie\in\{\leq,\geq\}$ and $\theta$ is a user-specified probability
threshold. The number of simulations required to decide the test is
typically fewer than (\ref{eq:NChernoff}) but is dependent on how
close $\theta$ is to the true probability. The number is therefore
not known in advance. To evaluate $\mathrm{P}(\omega\models\varphi)\bowtie\theta$,
the SPRT constructs hypotheses $H{}_{0}:\mathrm{P}(\omega\models\varphi)\geq p_{0}$
and $H{}_{1}:\mathrm{P}(\omega\models\varphi)\leq p_{1}$, where $p_{0}=\theta+\varepsilon$
and $p_{1}=\theta-\varepsilon$ for some user-defined interval specified
by $\varepsilon$ \cite{Wald1945}. The SPRT also requires parameters
$\alpha$ and $\beta$ to specify, respectively, the maximum acceptable
probabilities of incorrectly rejecting a true $H_{0}$ and incorrectly
accepting a false $H_{0}$. To choose between $H_{0}$ and $H_{1}$,
the SPRT defines the probability ratio 
\[
\mathit{ratio}=\prod_{i=1}^{n}\frac{(p^{1})^{\mathbf{1}(\omega_{i}\models\varphi)}(1-p^{1})^{\mathbf{1}(\omega_{i}\not\models\varphi)}}{(p^{0})^{\mathbf{1}(\omega_{i}\models\varphi)}(1-p^{0})^{\mathbf{1}(\omega_{i}\not\models\varphi)}},
\]
where $n$ is the number of simulation traces $\omega_{i}$, $i\in\{1,\dots$,
$n\}$, generated so far. The test proceeds by performing a simulation
and calculating $\mathit{ratio}$ until one of two conditions is satisfied:
$H_{1}$ is accepted if $\mathit{ratio}\geq(1-\beta)/\alpha$ and
$H_{0}$ is accepted if $\mathit{ratio}\leq\beta/(1-\alpha)$.

Parallelisation of SMC is conceptually simple with lightweight algorithms,
but balancing the simulation load on unreliable or heterogeneous computing
devices must be achieved without introducing a ``selection bias''.
The problem arises because simulation traces that satisfy a property
will, in general, take a different time to generate than those which
do not. If the SMC task is divided among a number of clients of different
speed or reliability, a naive balancing approach will be biased in
favour of results that are generated quickly. To overcome this phenomenon,
\textsc{Plasma} adopts the load balancing algorithm proposed in \cite{YounesPhD}.
\textsc{Plasma}'s GUI facilitates easy parallelisation on ad hoc networked
computers or on dedicated grids and clusters. The server application
(an instance of \textsc{Plasma}) starts the job and waits to be contacted
by available clients (instances of \textsc{Plasma} Service). Our estimation
experiments in Section \ref{sec:experiments} were distributed on
the \textsc{Igrida} computing grid%
\footnote{igrida.gforge.inria.fr%
}.

\section{Related Work\label{sec:related}}

The classic algorithms to solve MDPs are \emph{policy iteration} and
\emph{value iteration} \cite{Puterman1994}. Model checking algorithms
for MDPs may use value iteration applied to probabilities \cite[Ch. 10]{BaierKatoen2008}
or solve the same problem using linear programming \cite{BiancoDeAlfaro1995}.
All consider \emph{history-dependent} schedulers. The principal challenge
of finding optimal schedulers is what has been described as the `curse
of dimensionality' \cite{Bellman1957} and the `state explosion
problem' \cite{ClarkeEmersonAllenSifakis2009}. In essence, these
two terms refer to fact that the number of states of a system increases
exponentially with respect to the number of interacting components
and state variables. This phenomenon has motivated the design of lightweight
sampling algorithms that find `near optimal' schedulers to optimise
rewards in discounted MDPs, but the standard model checking problem
of finding extremal probabilities in non-discounted MDPs is significantly
more challenging. Since nondeterministic and probabilistic choices
are interleaved in an MDP, schedulers are typically of the same order
of complexity as the system as a whole and may be infinite. As a result,
previous SMC algorithms for MDPs have considered only memoryless schedulers
or have other limitations.

The Kearns algorithm \cite{KearnsMansourNg2002} is the classic `sparse
sampling algorithm' for large, infinite horizon, discounted MDPs.
It constructs a `near optimal' scheduler by approximating the best
action from a current state, using a stochastic depth-first search.
Importantly, optimality is with respect to discounted rewards, not
probability. The algorithm can work with large, potentially infinite
state MDPs because it explores a probabilistically bounded search
space. This, however, is exponential in the discount. To find the
action with the greatest expected reward in the current state of a
trace, the algorithm recursively estimates the rewards of successor
states, up to some maximum depth defined by the discount and desired
error. Actions are enumerated while probabilistic choices are explored
by sampling, with the number of samples set as a parameter. The discount
guarantees that the algorithm eventually converges. The stopping criterion
is when successive estimates differ by less than some error threshold.
Since the actions of a state are re-evaluated every time the state
is visited (because actions are history-dependent), the performance
of the Kearns algorithm is critically dependent on its parameters.

There have been several recent attempts to apply SMC to nondeterministic
models \cite{Bogdoll2011,LassaignePeyronnet2012,Henriques-et-al2012,HartmannsTimmer2013,LegaySedwardsTraonouez2014}.
In \cite{Bogdoll2011,HartmannsTimmer2013} the authors present on-the-fly
algorithms to remove `spurious' nondeterminism, so that standard
SMC may be used. This approach is limited to the class of models whose
nondeterminism does not affect the resulting probability of a property.
The algorithms therefore do not attempt to address the standard MDP
model checking problems related to finding optimal schedulers.

In \cite{LassaignePeyronnet2012} the authors first find a memoryless
scheduler that is near optimal with respect to a reward scheme and
discount, using an adaptation of the Kearns algorithm. This induces
a Markov chain whose properties may be verified with standard SMC.
By storing and re-using the choices in visited states, the algorithm
improves on the performance of the Kearns algorithm, but is thus limited
to tractable memoryless schedulers. The near optimality of the induced
Markov chain is with respect to rewards, not probability, hence \cite{LassaignePeyronnet2012}
does not address the standard model checking problems of MDPs.

In \cite{Henriques-et-al2012} the authors present an SMC algorithm
to decide whether there exists a memoryless scheduler for a given
MDP, such that the probability of a property is above a given threshold.
The algorithm has an inner loop that generates candidate schedulers
by iteratively improving a probabilistic scheduler, according to sample
traces that satisfy the property. The algorithm is limited to memoryless
schedulers because the improvement process learns by counting state-action
pairs. The outer loop tests the candidate scheduler against the hypothesis
using SMC and is iterated until an example is found or sufficient
attempts have been made. The approach has several problems. The inner
loop does not in general converge to the true optimum (the number
of state-actions does not actually indicate scheduler probability),
but is sometimes successful because the outer loop randomly explores
local maxima. This makes the number of samples used by the inner loop
critical: too many may reduce the randomness of the outer loop's exploration
and thus significantly reduce the probability of finding examples.
A further problem is that the repeated hypothesis tests of the outer
loop will eventually produce erroneous results.

The present work builds on the elements of lightweight verification
for MDPs introduced in \cite{LegaySedwardsTraonouez2014}. In \cite{LegaySedwardsTraonouez2014}
the authors use an incremental hash function and a pseudo-random number
generator to define history-dependent schedulers using only $\mathcal{O}(1)$
memory. This allows the schedulers to be selected at random and tested
individually, thus facilitating Monte Carlo algorithms that are indifferent
to the size of the sample space. The full details of these techniques
are described in Section \ref{sec:seeds}.

\section{Lightweight Verification of MDPs\label{sec:seeds}}

In this section we recall the elemental sampling techniques of \cite{LegaySedwardsTraonouez2014}.

Storing schedulers as explicit mappings does not scale, so we represent
schedulers using uniform pseudo-random number generators (PRNG) that
are initialised by a \emph{seed} and iterated to generate the next
pseudo-random value. In general, such PRNGs aim to ensure that arbitrary
subsets of sequences of iterates are uniformly distributed and that
consecutive iterates are statistically independent. PRNGs are commonly
used to implement the uniform probabilistic scheduler, which chooses
actions uniformly at random and thus explores all possible combinations
of nondeterministic choices. Executing such an implementation twice
with the same seed will produce identical traces. Executing the implementation
with a different seed will produce an unrelated set of choices: individual
schedulers cannot be identified, so it is not possible to estimate
the probability of a property under a specific scheduler. We use a
PRNG to resolve nondeterministic choices, but not to make those choices
probabilistically. We use the PRNG to range over the possible choices,
such that repeated scheduler samplings will eventually consider all
possible sequences of actions. We also rely on the fact that the seed
of a PRNG uniquely defines the sequence of pseudo-random values. Hence,
in contrast to the uniform probabilistic scheduler, actions are consistent
between simulations.

An apparently plausible solution is to use independent PRNGs to resolve
nondeterministic and probabilistic choices. It is then possible to
generate multiple probabilistic simulation traces per scheduler by
keeping the seed of the PRNG for nondeterministic choices fixed while
choosing random seeds for a separate PRNG for probabilistic choices.
Unfortunately, the schedulers generated by this approach do not span
the full range of general or even memoryless schedulers. Since the
sequence of iterates from the PRNG used for nondeterministic choices
will be the same for all instantiations of the PRNG used for probabilistic
choices, the $i^{\mathrm{th}}$ iterate of the PRNG for nondeterministic
choices will always be the same, regardless of the state arrived at
by the previous probabilistic choices. The $i^{\mathrm{th}}$ chosen
action can be neither state nor trace dependent, as required by memoryless
($\mathfrak{M}$) and history-dependent ($\mathfrak{S}$) schedulers,
respectively.

\subsection{General Schedulers Using Hash Functions\label{sec:hash}}

We therefore construct a per-step PRNG seed that is a \emph{hash}
of the integer identifying a specific scheduler concatenated with
an integer representing the sequence of states up to the present.

We assume that a state of an MDP is an assignment of values to a vector
of system variables $v_{i},i\in\{1,\dots,n\}$. Each $v_{i}$ is represented
by a number of bits $b_{i}$, typically corresponding to a primitive
data type (\emph{int}, \emph{float}, \emph{double}, etc.). The state
can thus be represented by the concatenation of the bits of the system
variables, such that a sequence of states may be represented by the
concatenation of the bits of all the states. Without loss of generality,
we interpret such a sequence of states as an integer of $\sum_{i=1}^{n}b_{i}$
bits, denoted $\overline{s}$, and refer to this in general as the
\emph{trace vector}. A scheduler is denoted by an integer $\sigma$,
which is concatenated to $\overline{s}$ (denoted $\sigma:\overline{s}$)
to uniquely identify a trace and a scheduler. Our approach is to generate
a hash code $h=\mathcal{H}(\sigma:\overline{s})$ and to use $h$
as the seed of a PRNG that resolves the next nondeterministic choice.

The hash function $\mathcal{H}$ thus maps $\sigma:\overline{s}$
to a seed that is deterministically dependent on the trace and the
scheduler. The PRNG maps the seed to a value that is uniformly distributed
but nevertheless deterministically dependent on the trace and the
scheduler. In this way we approximate the schedulers functions $\mathfrak{S}$
and $\mathfrak{M}$ described in Section \ref{sec:prelims}. Importantly,
the technique only relies on the standard properties of hash functions
and PRNGs. Algorithm \ref{alg:simulate} is the basic simulation function
used by our algorithms.

\begin{algorithm}
\KwIn{\\\Indp

$\mathcal{M}$: an MDP with initial state $s_{0}$

$\varphi$: a property

$\sigma$: an integer identifying a scheduler

}\KwOut{\\\Indp

$\omega$: a simulation trace

}\BlankLine

Let $\mathcal{U}_{\mathrm{prob}},\mathcal{U}_{\mathrm{nondet}}$ be
uniform PRNGs with respective samples $r_{\mathrm{pr}},r_{\mathrm{\mathrm{nd}}}$

Let $\mathcal{H}$ be a hash function

Let $s$ denote a state, initialised $s\leftarrow s_{0}$

Let $\omega$ denote a trace, initialised $\omega\leftarrow s$

Let $\overline{s}$ be the trace vector, initially empty

Set seed of $\mathcal{U}_{\mathrm{prob}}$ randomly

\While{$\omega\models\varphi$ is not decided}{

$\overline{s}\leftarrow\overline{s}:s$ 

Set seed of $\mathcal{U}_{\mathrm{nondet}}$ to $\mathcal{H}(\sigma:\overline{s})$

Iterate $\mathcal{U}_{\mathrm{nondet}}$ to generate $r_{\mathrm{nd}}$
and use to resolve nondeterministic choice

Iterate $\mathcal{U}_{\mathrm{prob}}$ to generate $r_{\mathrm{pr}}$
and use to resolve probabilistic choice

Set $s$ to the next state

$\omega\leftarrow\omega:s$

}

\caption{Simulate\label{alg:simulate}}
\end{algorithm}

\subsection{An Efficient Iterative Hash Function\label{sec:iterative}}

To implement our approach, we use an efficient hash function that
constructs seeds incrementally. The function is based on modular division
\cite[Ch. 6]{Knuth1998}, such that $h=(\sigma:\overline{s})\bmod m$,
where $m$ is a suitably large prime.

Since $\overline{s}$ is a concatenation of states, it is usually
very much larger than the maximum size of integers supported as primitive
data types. Hence, to generate $h$ we use Horner's method \cite{Horner1819}\cite[Ch. 4]{Knuth1998}:
we set $h_{0}=\sigma$ and find $h\equiv h_{n}$ ($n$ as in Section
\ref{sec:hash}) by iterating the recurrence relation 
\begin{equation}
h_{i}=(h_{i-1}2^{b_{i}}+v_{i})\bmod m.\label{eq:horner}
\end{equation}

The size of $m$ defines the maximum number of different hash codes.
The precise value of $m$ controls how the hash codes are distributed.
To avoid collisions, a simple heuristic is that $m$ should be a large
prime not close to a power of 2 \cite[Ch. 11]{CormenLeiersonRivestStein2009}.
The number of schedulers is typically much larger than the number
of possible hash codes, hence collisions are theoretically inevitable.
This means that not all possible schedulers are realisable with a
given hash function and PRNG. Since our chosen hash function and PRNG
are drawn from respective families of hash functions and PRNGs that
potentially span all schedulers, the problem of collisions can conceivably
be addressed by also choosing the hash function and PRNG at random.
A scheduler would then be defined by its label, its hash function
and its PRNG. We do not implement this idea here to avoid unnecessary
complication and because collisions are not the principal limitation.
There are typically many orders of magnitude more seeds than we can
test, hence the problem of finding the best available scheduler supersedes
the problem that the best available scheduler may not be optimal.
We anticipate that our proposed solutions to accelerate convergence
(property-focused scheduler space and piecewise construction of schedulers)
will effectively bypass the collision problem. 

In practical implementations it is an advantage to perform calculations
using primitive data types that are native to the computational platform,
so the sum in (\ref{eq:horner}) should always be less than or equal
to the maximum permissible value. To achieve this, given $x,y,m\in\mathbb{N}$,
we note the following congruences:
\begin{eqnarray}
(x+y)\bmod m & \equiv & (x\bmod m+y\bmod m)\bmod m\label{eq:modadd}\\
(xy)\bmod m & \equiv & ((x\bmod m)(y\bmod m))\bmod m\label{eq:modmul}
\end{eqnarray}
The addition in (\ref{eq:horner}) can thus be re-written in the form
of (\ref{eq:modadd}), such that each term has a maximum value of
$m-1$:
\begin{equation}
h_{i}=((h_{i-1}2^{b_{i}})\bmod m+(v_{i})\bmod m)\bmod m\label{eq:horner2}
\end{equation}

To prevent overflow, $m$ must be no greater than half the maximum
possible integer. Re-writing the first term of (\ref{eq:horner2})
in the form of (\ref{eq:modmul}), we see that before taking the modulus
it will have a maximum value of $(m-1)^{2}$, which will exceed the
maximum possible integer. To avoid this, we take advantage of the
fact that $h_{i-1}$ is multiplied by a power of 2 and that $m$ has
been chosen to prevent overflow with addition. We thus apply the following
recurrence relation:

\begin{eqnarray}
(h_{i-1}2^{j})\bmod m & = & (h_{i-1}2^{j-1})\bmod m\nonumber \\
 & + & (h_{i-1}2^{j-1})\bmod m\label{eq:shiftmod}
\end{eqnarray}

Equation (\ref{eq:shiftmod}) allows our hash function to be implemented
using efficient native arithmetic. Moreover, we infer from (\ref{eq:horner})
that to find the hash code corresponding to the current state in a
trace, we need only know the current state and the hash code from
the previous step. When considering memoryless schedulers we need
only know the current state.

\subsection{Hypothesis Testing Multiple Schedulers\label{sub:wald}}

To decide whether there exists a scheduler such that $\mathrm{P}(\omega\models\varphi)\bowtie p$,
we apply the SPRT to multiple (randomly chosen) schedulers. Since
the probability of error with the SPRT applied to multiple hypotheses
is cumulative, we consider the probability of no errors in any of
$M$ tests. Hence, in order to ensure overall error probabilities
$\alpha$ and $\beta$, we adopt $\alpha_{M}=1-\sqrt[M]{1-\alpha}$
and $\beta_{M}=1-\sqrt[M]{1-\beta}$ in our stopping conditions. $H_{1}$
is accepted if $\mathit{ratio}\geq(1-\beta_{M})/\alpha_{M}$ and $H_{0}$
is accepted if $\mathit{ratio}\leq\beta_{M}/(1-\alpha_{M})$. Algorithm
\ref{alg:SPRT} demonstrates the sequential hypothesis test for multiple
schedulers. If the algorithm finds an example, the hypothesis is true
with at least the specified confidence.

\begin{algorithm}[h]
\KwIn{\\\Indp

$\mathcal{M},\varphi$: the MDP and property of interest

$H\in\{H_{0},H_{1}\}$: the hypothesis with interval $\theta\pm\varepsilon$

$\alpha,\beta$: the desired error probabilities of $H$

$M$: the maximum number of schedulers to test

}\KwOut{The result of the hypothesis test}\BlankLine

Let $p^{0}=\theta+\varepsilon$ and $p^{1}=\theta-\varepsilon$ be
the bounds of $H$

Let $\alpha_{M}=1-\sqrt[M]{1-\alpha}$ and $\beta_{M}=1-\sqrt[M]{1-\beta}$

Let $A=(1-\beta_{M})/\alpha_{M}$ and $B=\beta_{M}/(1-\alpha_{M})$

Let $\mathcal{U}_{\mathrm{seed}}$ be a uniform PRNG and $\sigma$
be its sample

\For{$i\in\{1,\dots,M\}$ while $H$ is not accepted}{

Iterate $\mathcal{U}_{\mathrm{seed}}$ to generate $\sigma_{i}$

Let $\mathit{ratio}=1$

\While{$\mathit{ratio}>A\wedge\mathit{ratio}<B$}{

$\omega\leftarrow\mathrm{Simulate}(\mathcal{M},\varphi,\sigma_{i})$

$\mathit{ratio}\leftarrow\frac{(p^{1})^{\mathbf{1}(\omega\models\varphi)}(1-p^{1})^{\mathbf{1}(\omega\not\models\varphi)}}{(p^{0})^{\mathbf{1}(\omega\models\varphi)}(1-p^{0})^{\mathbf{1}(\omega\not\models\varphi)}}\mathit{ratio}$

}\If{$\mathit{ratio}\leq A\wedge H=H_{0}\vee\mathit{ratio}\geq B\wedge H=H_{1}$}{accept
$H$

}}

\caption{SPRT for multiple schedulers\label{alg:SPRT}}
\end{algorithm}

\subsection{Estimating Multiple Schedulers\label{sub:chernoff}}

We consider the strategy of sampling $M$ schedulers to estimate the
optimum probability. We thus generate $M$ estimates $\{\hat{p}_{1},\dots,\hat{p}_{M}\}$,
corresponding to true values $\{p_{1},$ $\dots,p_{M}\}$, and take
either the maximum ($\hat{p}_{\max}$) or minimum ($\hat{p}_{\min}$),
as required. To overcome the cumulative probability of error with
the standard Chernoff bound, we specify that \emph{all} estimates
$\hat{p}_{i}$ must be within $\varepsilon$ of their respective true
values $p_{i}$, ensuring that any $\hat{p}_{\min},\hat{p}_{\max}\in\{\hat{p}_{1},\dots,\hat{p}_{M}\}$
are within $\varepsilon$ of their true value. Given that all estimates
$\hat{p}_{i}$ are statistically independent, the probability that
all estimates are less than their upper bound is expressed by $\mathrm{P}(\bigwedge_{i=1}^{M}\hat{p}_{i}-p_{i}\leq\varepsilon)\geq(1-e^{-2N\varepsilon^{2}})^{M}$.
Hence, $\mathrm{P}(\bigvee_{i=1}^{M}\hat{p}_{i}-p_{i}\geq\varepsilon)\leq1-(1-e^{-2N\varepsilon^{2}})^{M}$,
giving $N=\left\lceil -\ln\left(1-\sqrt[M]{1-\delta}\right)/(2\varepsilon^{2})\right\rceil $
for user-specified parameters $M$, $\varepsilon$ and $\delta$.
This ensures that $\mathrm{P}(p_{\min}-\hat{p}_{\min}\geq\varepsilon)\leq\delta$
and $\mathrm{P}(\hat{p}_{\max}-p_{\max}\geq\varepsilon)\leq\delta$.
To ensure the usual stronger conditions that $\mathrm{P}(\mid p_{\max}-\hat{p}_{\max}\mid\geq\varepsilon)\leq\delta$
and $\mathrm{P}(\mid p_{\min}-\hat{p}_{\min}\mid\geq\varepsilon)\leq\delta$,
we have

\begin{equation}
N=\left\lceil \left(\ln2-\ln\left(1-\sqrt[M]{1-\delta}\right)\right)/(2\varepsilon^{2})\right\rceil .\label{eq:NMChernoff}
\end{equation}

$N$ scales logarithmically with $M$, making it tractable to consider
many schedulers. In the case of $M=1$, (\ref{eq:NMChernoff}) degenerates
to (\ref{eq:NChernoff}). Note, however, that the confidence expressed
by (\ref{eq:NMChernoff}) is with respect to the sampled set, not
with respect to the true extrema. Algorithm \ref{alg:chernoff} is
the resulting extremal probability estimation algorithm for multiple
schedulers.

\begin{algorithm}[h]
\KwIn{\\\Indp

$\mathcal{M},\varphi$: the MDP and property of interest

$\varepsilon,\delta$: the required Chernoff bound

$M$: the number of schedulers to test

}\KwOut{Extremal estimates $\hat{p}_{\min}$ and $\hat{p}_{\max}$}\BlankLine

Let $N=\left\lceil \ln(2/(1-\sqrt[M]{1-\delta}\,))/(2\varepsilon^{2})\right\rceil $
be the no. of simulations per scheduler

Let $\mathcal{U}_{\mathrm{seed}}$ be a uniform PRNG and $\sigma$
its sample

Initialise $\hat{p}_{\min}\leftarrow1$ and $\hat{p}_{\max}\leftarrow0$

Set seed of $\mathcal{U}_{\mathrm{seed}}$ randomly

\For{$i\in\{1,\dots,M\}$}{

Iterate $\mathcal{U}_{\mathrm{seed}}$ to generate $\sigma{}_{i}$

Let $\mathit{truecount}=0$ be the initial number of traces that satisfy
$\varphi$

\For{$j\in\{1,\dots,N\}$}{

$\omega_{j}\leftarrow\mathrm{Simulate}(\mathcal{M},\varphi,\sigma_{i})$

$\mathit{truecount}\leftarrow\mathit{truecount}+\mathbf{1}(\omega_{j}\models\varphi)$

}

Let $\hat{p}_{i}=\mathit{truecount}/N$

\If{$\hat{p}_{\max}<\hat{p}_{i}$}{$\hat{p}_{\max}=\hat{p}_{i}$

}\If{$\hat{p}_{i}>0\wedge\hat{p}_{\min}>\hat{p}_{i}$}{$\hat{p}_{\min}=\hat{p}_{i}$

}}

\If{$\hat{p}_{\max}=0$}{No schedulers were found to satisfy $\varphi$

}

\caption{Estimation with multiple schedulers\label{alg:chernoff}}
\end{algorithm}

\begin{figure}
\centering{}\includegraphics[height=0.5\columnwidth]{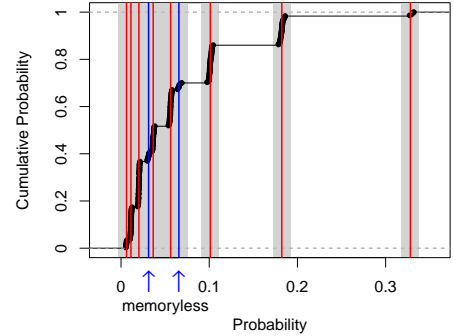}\caption{Empirical cumulative distribution of estimates from Algorithm \ref{alg:chernoff}
applied to MDP of Fig. \ref{fig:historydependent}.\label{fig:schedulers}}
\end{figure}

Figure \ref{fig:schedulers} shows the empirical cumulative distribution
of schedulers generated by Algorithm \ref{alg:chernoff} applied to
the MDP of Fig. \ref{fig:historydependent}, using $p_{1}=0.9$, $p_{2}=0.5$,
$\varphi=\mathbf{X}(\psi\wedge\mathbf{XG}^{4}\neg\psi)$, $\varepsilon=0.01$,
$\delta=0.01$ and $M=300$. The vertical red and blue lines mark
the true probabilities of $\varphi$ under each of the history-dependent
and memoryless schedulers, respectively. The grey rectangles show
the $\pm\varepsilon$ error bounds, relative to the true probabilities.
There are multiple estimates per scheduler, but all estimates are
within their respective confidence bounds.

\section{Smart Sampling\label{sec:smart}}

The simple sampling strategies used by Algorithms \ref{alg:SPRT}
and \ref{alg:chernoff} have the disadvantage that they allocate equal
simulation budget to all schedulers, regardless of their merit. In
general, the problem we address has two independent components: the
rarity of near optimal schedulers and the rarity of the property under
a near optimal scheduler. We should allocate our simulation budget
accordingly and not waste budget on schedulers that are clearly not
optimal.

Motivated by the above, our smart estimation algorithm comprises three
stages: (\emph{i}) an initial undirected sampling experiment to discover
the nature of the problem, (\emph{ii}) a targeted sampling experiment
to generate a candidate set of schedulers with high probability of
containing an optimal scheduler and (\emph{iii}) iterative refinement
of the candidates to estimate the probability of the best scheduler
with specified confidence. By excluding the schedulers with the worst
estimated probabilities and re-allocating their simulation budget
to the schedulers that remain, at each iterative step of stage (\emph{iii})
the number of schedulers reduces while the confidence of their estimates
increases. With a suitable choice of per-iteration budget, the algorithm
is guaranteed to terminate.

In the following subsection we develop the theoretical basis of stage
(\emph{ii}).

\subsection{Maximising the Probability of Seeing a Good Scheduler\label{sec:maximising}}

In what follows we assume the existence of an MDP and a property $\varphi$
whose probability we wish to maximise by choosing a suitable scheduler
from the set $\mathfrak{S}$. Let $\mathcal{P}:\mathfrak{S}\rightarrow[0,1]$
be a function mapping schedulers to their probability of satisfying
$\varphi$ and let $p_{\max}=\max_{\sigma\in\mathfrak{S}}(\mathcal{P}(s))$.
For the sake of exposition, we consider the problem of finding a scheduler
that maximises the probability of satisfying $\varphi$ and define
a ``good'' (near optimal) scheduler to be one in the set $\mathfrak{S}_{g}=\{\sigma\in\mathfrak{S}\mid\mathcal{P}(\sigma)\geq p_{\max}-\varepsilon\}$
for some $\varepsilon\in(0,p_{\max}]$. Intuitively, a good scheduler
is one whose probability of satisfying $\varphi$ is within $\varepsilon$
of $p_{\max}$, noting that we may similarly define a good scheduler
to be one within $\varepsilon$ of $p_{\min}=\min_{\sigma\in\mathfrak{S}}(\mathcal{P}(\sigma))$,
or to be in any other subset of $\mathfrak{S}$. In particular, to
address reward-based MDP optimisations, a good scheduler could be
defined to be the subset of $\mathfrak{S}$ that is near optimal with
respect to a reward scheme. The notion of a ``best'' scheduler follows
intuitively from the definition of a good scheduler.

If we sample uniformly from $\mathfrak{S}$, the probability of finding
a good scheduler is $p_{g}=|\mathfrak{S}_{g}|/|\mathfrak{S}|$. The
average probability of a good scheduler is $p_{\overline{g}}=\sum_{\sigma\in\mathfrak{S}_{g}}\mathcal{P}(\sigma)/|\mathfrak{S}_{g}|$.
If we select $M$ schedulers uniformly at random and verify each with
$N$ simulations, the expected number of traces that satisfy $\varphi$
using a good scheduler is thus $Mp_{g}Np_{\overline{g}}$. The probability
of seeing a trace that satisfies $\varphi$ using a good scheduler
is the cumulative probability 
\begin{equation}
(1-(1-p_{g})^{M})(1-(1-p_{\overline{g}})^{N}).\label{eq:probgoodsched}
\end{equation}
Hence, for a given simulation budget $N_{\max}=NM$, to implement
stage (\emph{ii}) the idea is to choose $N$ and $M$ to maximise
(\ref{eq:probgoodsched}) and keep any scheduler that produces at
least one trace that satisfies $\varphi$. Since, a priori, we are
generally unaware of even the magnitudes of $p_{g}$ and $p_{\overline{g}}$,
stage ($i$) is necessarily uninformed and we set $N=M=\lceil\sqrt{N_{\max}}\rceil$.
The results of stage (\emph{i}) allow us to estimate $p_{g}$ and
$p_{\overline{g}}$ (see Fig. \ref{fig:ECDFconvergence}) and thus
maximise (\ref{eq:probgoodsched}). This may be done numerically,
but we have found the heuristic $N=\lceil1/p_{\overline{g}}\rceil$
to be near optimal in all but extreme cases.

\subsection{Smart Estimation\label{sub:smartest}}

Algorithm \ref{alg:smartest} is our smart estimation algorithm to
find schedulers that maximise the probability of a property. The algorithm
to find minimising schedulers is similar. Lines \ref{alg:stageistart}
to \ref{alg:stageiend} implement stage (\emph{i}), lines \ref{alg:stageiistart}
to \ref{alg:stageiiend} implement stage (\emph{ii}) and lines \ref{alg:stageiiistart}
to \ref{alg:stageiiiend} implement stage (\emph{iii}). Note that
the algorithm distinguishes $p_{\max}$ (the notional true maximum
probability), $p_{\overline{\max}}$ (the true probability of the
best candidate scheduler found) and $\hat{p}_{\overline{\max}}$ (the
estimated probability of the best candidate scheduler).

The per-iteration simulation budget $N_{\max}$ must be greater than
the number needed by the standard Chernoff bound (\ref{eq:NChernoff}),
to ensure that there will be sufficient simulations to guarantee the
specified confidence if the algorithm refines the candidate set to
a single scheduler. Typically, the per-iteration budget will be greater
than this, such that the required confidence is reached before refining
the set of schedulers to a single element. Under these circumstances
the confidence is judged according to the Chernoff bound for multiple
estimates (\ref{eq:NMChernoff}). In addition, lines \ref{alg:quitiiistart}
to \ref{alg:quitiiiend} allow the algorithm to quit as soon as the
minimum number of simulations is reached.

Algorithm \ref{alg:smartest} may be further optimised by re-using
the simulation results from previous iterations of stage (\emph{iii}).
The contribution is small, however, because confidence decreases exponentially
with the age (in terms of iterations) of the results.

\begin{algorithm}[h]
\KwIn{\\\Indp

$\mathcal{M}$: an MDP 

$\varphi$: a property

$\epsilon,\delta$: the required Chernoff bound

$N_{\max}>\ln(2/\delta)/(2\epsilon^{2})$: the per-iteration budget

}\KwOut{$\hat{p}_{\overline{\max}}\approx p_{\max}$, where $p_{\overline{\max}}\approx p_{\max}$
and $\mathrm{P}(|p_{\overline{\max}}-\hat{p}_{\overline{\max}}|\geq\epsilon)\leq\delta$}\BlankLine

$N\leftarrow\lceil\sqrt{N_{\max}}\rceil$; $M\leftarrow\lceil\sqrt{N_{\max}}\rceil$\label{alg:stageistart}

$S\leftarrow\{M\textnormal{ seeds chosen uniformly at random}\}$ 

$\forall\sigma\in S,\forall i\in\{1,\dots,N\}:\omega_{i}^{\sigma}\leftarrow\mathrm{Simulate}(\mathcal{M},\varphi,\sigma)$ 

$R:S\rightarrow\mathbb{N}$ maps scheduler seeds to number of traces
satisfying $\varphi$:

\nonl$\quad R\leftarrow\{(\sigma,n)\mid\sigma\in S\wedge\mathbb{N}\ni n=\sum_{i=1}^{N}\mathbf{1}(\omega_{i}^{\sigma}\models\varphi)\}$

$\hat{p}_{\overline{\max}}\leftarrow\max_{\sigma\in S}(R(\sigma)/N)$\label{alg:stageiend}

$N\leftarrow\lceil1/\hat{p}_{\overline{\max}}\rceil$, $M\leftarrow\lceil N_{\max}\,\hat{p}_{\overline{\max}}\rceil$\label{alg:stageiistart}

$S\leftarrow\{M\textnormal{ seeds chosen uniformly at random}\}$

$\forall\sigma\in S,\forall i\in\{1,\dots,N\}:\omega_{i}^{\sigma}\leftarrow\mathrm{Simulate}(\mathcal{M},\varphi,\sigma)$ 

$R\leftarrow\{(\sigma,n)\mid\sigma\in S\wedge\mathbb{N}\ni n=\sum_{i=1}^{N}\mathbf{1}(\omega_{i}^{\sigma}\models\varphi)\}$

$S\leftarrow\{\sigma\in S\mid R(\sigma)>0\}$\label{alg:stageiiend}

$\forall\sigma\in S,\; R(\sigma)\leftarrow0$; $i\leftarrow0$; $\mathit{conf}\leftarrow1$\label{alg:stageiiistart}

\While{$\mathit{conf}>\delta\wedge S\neq\emptyset$}{

$i\leftarrow i+1$

$M_{i}\leftarrow|S|$

$N_{i}\leftarrow0$

\While{$\mathit{conf}>\delta\wedge N_{i}<\lceil N_{\max}/M_{i}\rceil$}{\label{alg:quitiiistart}

$N_{i}\leftarrow N_{i}+1$

$\mathit{conf}\leftarrow1-(1-e^{-2\epsilon^{2}N_{i}})^{M_{i}}$

$\forall\sigma\in S:\omega_{N_{i}}^{\sigma}\leftarrow\mathrm{Simulate}(\mathcal{M},\varphi,\sigma)$\label{alg:quitiiiend}

}

$R\leftarrow\{(\sigma,n)\mid\sigma\in S\wedge\mathbb{N}\ni n=\sum_{j=1}^{N_{i}}\mathbf{1}(\omega_{j}^{\sigma}\models\varphi)\}$

$\hat{p}_{\overline{\max}}\leftarrow\max_{\sigma\in S}(R(\sigma)/N_{i})$

$R':\{1,\dots,|S|\}\rightarrow S$ is an injective function s.t.

\nonl\quad{}$\forall(n,\sigma),(n',\sigma')\in R',\; n>n'\implies R(\sigma)\geq R(\sigma')$

$S\leftarrow\{\sigma\in S\mid\sigma=R'(n)\wedge n\in\{\lfloor|S|/2\rfloor,\dots,|S|\}\}$

\label{alg:stageiiiend}

}

\caption{Smart Estimating\label{alg:smartest}}
\end{algorithm}

\subsection{Smart Hypothesis Testing\label{sec:smarthyp}}

We wish to test the hypothesis that there exists a scheduler such
that property $\varphi$ has probability $\bowtie\theta$, where $\bowtie\in\{\geq,\leq\}$.
Two advantages of sequential hypothesis testing are that it is not
necessary to estimate the actual probability to know if an hypothesis
is satisfied, and the easier the hypothesis is to satisfy, the quicker
it is to get a result. Algorithm \ref{alg:smarthyp} maintains these
advantages and uses smart sampling to improve on the performance of
Algorithm \ref{alg:SPRT}. For the purposes of exposition, Algorithm
\ref{alg:smarthyp} tests $H_{0}$, as described in Section \ref{sec:smc}.
The algorithm to test $H_{1}$ is similar.

A sub-optimal approach would be to use Algorithm \ref{alg:smartest}
to refine a set of schedulers until one is found whose estimate satisfies
the hypothesis with confidence according to a Chernoff bound. Our
approach is to exploit the fact that the \emph{average} estimate at
each iteration of Algorithm \ref{alg:smartest} is known with high
confidence, i.e., confidence given by the total simulation budget.
This follows directly from the result of \cite{Chernoff1952}, where
the bound is specified for a sum of arbitrary random variables, not
necessarily with identical expectations. By similar arguments based
on \cite{Wald1945}, it follows that the sequential probability ratio
test may also be applied to the sum of results produced during the
course of an iteration of Algorithm \ref{alg:smartest}. Moreover,
it is possible to test each scheduler with respect to its individual
results and the current number of schedulers, according to the bound
given in Section \ref{sub:wald}.

Hence, if the ``average scheduler'' or an individual scheduler ever
satisfies the hypothesis (lines \ref{alg:hypsatstart} and \ref{alg:hypsatend}),
the algorithm immediately terminates and reports that the hypothesis
is satisfied with the specified confidence. If the ``best'' scheduler
ever individually falsifies the hypothesis (lines \ref{alg:hypfalsestart}
and \ref{alghypfalseend}), the algorithm also terminates and reports
the result. Note that this outcome does not imply that there is no
scheduler that will satisfy the hypothesis, only that no scheduler
was found with the given budget. Finally, if the algorithm refines
the initial set of schedulers to a single instance and the hypothesis
was neither satisfied nor falsified, an inconclusive result is reported
(line \ref{alg:hypincinclusive}).

We implement one further important optimisation. We use the threshold
probability $\theta$ to directly define the simulation budget to
generate the candidate set of schedulers, i.e. $N=\lceil1/\theta\rceil$,
$M=\lceil\theta N_{\max}\rceil$ (line \ref{alg:smarthypiistart}).
This is justified because we need only find schedulers whose probability
of satisfying $\varphi$ is greater than $\theta$. By setting $N=\lceil1/\theta\rceil$,
(\ref{eq:probgoodsched}) ensures that such schedulers, if they exist,
have high probability of being observed. The initial coarse exploration
used in Algorithm \ref{alg:smartest} is thus not necessary.

Algorithm \ref{alg:smarthyp} is our smart hypothesis testing algorithm.
Note that we do not set a precise minimum per-iteration simulation
budget because we expect the hypothesis to be decided with many fewer
simulations than would be required to estimate the probability. In
practice it is expedient to initially set a low per-iteration budget
(e.g., 1000) and repeat the algorithm with an increased budget (e.g.,
increased by an order of magnitude) if the previous test was inconclusive.

\begin{algorithm}
\KwIn{\\\Indp

$\mathcal{M}$: an MDP

$\varphi$: a property

$H_{0}:\mathrm{P}(\omega\models\varphi)\geq\theta\pm\varepsilon$
is the hypothesis

$\alpha,\beta$: the desired error probabilities of $H_{0}$

$N_{\max}$: the per-iteration simulation budget

}\KwOut{The result of the hypothesis test}\BlankLine

Let $p^{0}=\theta+\varepsilon$, $p^{1}=\theta-\varepsilon$

Let $A=(1-\beta)/\alpha$, $B=\beta/(1-\alpha)$

$N\leftarrow\lceil1/\theta\rceil$; $M\leftarrow\lceil\theta N_{\max}\rceil$\label{alg:smarthypiistart}

$S\leftarrow\{M\textnormal{ seeds chosen uniformly at random}\}$
\label{alg:smarthypiiend}

$\forall\sigma\in S,\forall i\in\{1,\dots,N\}:\omega_{i}^{\sigma}\leftarrow\mathrm{Simulate}(\mathcal{M},\varphi,\sigma)$

$R\leftarrow\{(\sigma,n)\mid\sigma\in S\wedge\mathbb{N}\ni n=\sum_{i=1}^{N}\mathbf{1}(\omega_{i}^{\sigma}\models\varphi)\}$

\If{$\frac{(p^{1})^{\sum R(\sigma)}(1-p^{1})^{N_{\max}-\sum R(\sigma)}}{(p^{0})^{\sum R(\sigma)}(1-p^{0})^{N_{\max}-\sum R(\sigma)}}\leq A$}{Accept
$H_{0}$ and quit}

$S\leftarrow\{\sigma\in S\mid R(\sigma)>0\}$, $M\leftarrow|S|+1$

\While{$M>1$}{

$M\leftarrow|S|$

Let $\alpha_{M}=1-\sqrt[M]{1-\alpha}$, $\beta_{M}=1-\sqrt[M]{1-\beta}$

Let $A_{M}=(1-\beta_{M})/\alpha_{M}$, $B_{M}=\beta_{M}/(1-\alpha_{M})$

Let $\mathit{ratio}=1$

\For{$\sigma_{i}\in S,i\in\{1,\dots,M\}$}{

Let $\mathit{ratio}_{i}=1$

\For{$j\in\{1,\dots,N\}$}{

$\omega\leftarrow\mathrm{Simulate}(\mathcal{M},\varphi,\sigma_{i})$

\If{$\omega\models\varphi$}{

$\mathit{ratio}\leftarrow\frac{p_{1}}{p_{0}}\mathit{ratio}$; $\mathit{ratio}_{i}\leftarrow\frac{p_{1}}{p_{0}}\mathit{ratio}_{i}$

}\Else{

$\mathit{ratio}\leftarrow\frac{1-p_{1}}{1-p_{0}}\mathit{ratio}$;
$\mathit{ratio}_{i}\leftarrow\frac{1-p_{1}}{1-p_{0}}\mathit{ratio}_{i}$

}\If{$\mathit{ratio}\leq A\vee\mathit{ratio}_{i}\leq A_{M}$\label{alg:hypsatstart}}{

Accept $H_{0}$ and quit\label{alg:hypsatend}}\If{$\mathit{ratio}_{i}\geq B_{M}$\label{alg:hypfalsestart}}{

Reject $H_{0}$ (given budget) and quit\label{alghypfalseend}}}}

$R':\{1,\dots,|S|\}\rightarrow S$ is an injective function s.t.

\nonl$\forall(n,\sigma),(n',\sigma')\in R',\; n>n'\implies R(\sigma)\geq R(\sigma')$

$S\leftarrow\{\sigma\in S\mid\sigma=R'(n)\wedge n\in\{\lfloor|S|/2\rfloor,\dots,|S|\}\}$

}Inconclusive result (given budget)\label{alg:hypincinclusive}

\caption{Smart Hypothesis Testing\label{alg:smarthyp}}

\end{algorithm}

\section{Case Studies\label{sec:experiments}}

To demonstrate the performance of smart sampling, we have implemented
Algorithms \ref{alg:smartest} and \ref{alg:smarthyp} in our statistical
model checking platform \textsc{Plasma} \cite{Boyer-et-al2013}. We
performed a number of experiments on standard models taken from the
numerical model checking literature, most of which can be found illustrated
on the \textsc{Prism} website%
\footnote{www.prismmodelchecker.org/casestudies/%
}. We found that all of our estimation experiments achieved their specified
Chernoff bounds ($\varepsilon=\delta=0.01$ in all cases) with a relatively
modest per-iteration simulation budget of $10^{5}$ simulations. The
actual number of simulation cores used for the estimation results
was subject to availability and varied between experiments. To facilitate
comparisons, in what follows we normalise all timings to be with with
respect to 64 cores. Typically, each data point was produced in a
few tens of seconds. Our hypothesis tests were performed on a single
machine, without distribution. Despite this, most experiments completed
in just a few seconds (some in fractions of a second), demonstrating
that our smart hypothesis testing algorithm is able to take advantage
of easy hypotheses.

\subsection{IEEE 802.11 Wireless LAN Protocol\label{sec:WLAN}}

\begin{figure}
\centering{}\includegraphics[width=0.55\columnwidth]{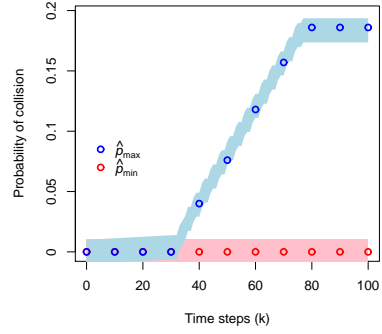}\caption{Estimated maximum and minimum probabilities of second collision in
WLAN protocol. Shaded regions denote true values $\pm0.01$. \label{fig:WLAN}}
\end{figure}

We consider a reachability property of the IEEE 802.11 Wireless LAN
(WLAN) protocol model of \cite{KwiatkowskaNormanSproston2002}. The
protocol aims to avoid ``collisions'' between devices sharing a
communication channel, by means of an exponential backoff procedure
when a collision is detected. We therefore estimate the probability
of the second collision at various time steps, using Algorithm \ref{alg:smartest}
with per-iteration budget of $10^{5}$ simulations. Fig. \ref{fig:WLAN}
illustrates the estimated maximum probabilities ($\hat{p}_{\max}$)
and minimum probabilities ($\hat{p}_{\min}$) for time steps $k\in\{0,10,\dots,100\}$.
The property is expressed as $\mathbf{F}^{k}col=2$. The shaded areas
indicate the true probabilities $\pm0.01$, the specified absolute
error bound using Chernoff bound $\varepsilon=\delta=0.01$. Our results
are clearly very close to the true values. Table \ref{tab:CSMA} gives
the results of hypothesis tests based on the same model using property
$\mathbf{F}^{100}col=2$. See Section \ref{sec:CSMA} for a description.

The results illustrated in Fig. \ref{fig:WLAN} refer to the same
property and confidence as the those shown in Fig. 4 of \cite{LegaySedwardsTraonouez2014}.
The total simulation cost to generate a point in Fig. \ref{fig:WLAN}
is $1.2\times10^{6}$ (12 iterations of $10^{5}$ simulations using
smart sampling), compared to a cost of $2.7\times10^{8}$ per point
in Fig. 4 of \cite{LegaySedwardsTraonouez2014} (4000 schedulers tested
with 67937 simulations using simple sampling). This demonstrates a
more than 200-fold improvement in performance.

\subsection{IEEE 802.3 CSMA/CD Protocol\label{sec:CSMA}}

The IEEE 802.3 CSMA/CD protocol is a wired network protocol that is
similar in operation to that of IEEE 802.11, but using collision detection
instead of collision avoidance. In Table \ref{tab:CSMA} we give the
results of applying Algorithm \ref{alg:smarthyp} to the IEEE 802.3
CSMA/CD protocol model of \cite{KwiatkowskaNormanParkerSprotson2006}.
The models and parameters are chosen to compare with results given
in Table III in \cite{Henriques-et-al2012}, hence we also give results
for hypothesis tests performed on the WLAN model used in Section \ref{sec:WLAN}.
In contrast to the results of \cite{Henriques-et-al2012}, our results
are produced on a single machine, with no parallelisation. There are
insufficient details given about the experimental conditions in \cite{Henriques-et-al2012}
to make a formal comparison (e.g., error probabilities of the hypothesis
tests and number of simulation cores), but it seems that the performance
of our algorithm is generally much better. We set $\alpha=\beta=\delta=0.01$,
which constitute a fairly tight bound, and note that, as expected,
the simulation times tend to increase as the threshold $\theta$ approaches
the true probability.

\begin{table}
\begin{centering}
{\smaller%
\begin{tabular}{|l|c|c|c|c|c|c|c|}
\hline 
\multirow{2}{*}{CSMA 3\,4} & $\theta$ & 0.5 & 0.8 & 0.85 & 0.86 & 0.9 & 0.95\tabularnewline
\cline{2-8} 
 & \emph{time} & 0.5 & 3.5 & 737 & {*} & 2.9 & 2.5\tabularnewline
\hline 
\hline 
\multirow{2}{*}{CSMA 3\,6} & $\theta$ & 0.3 & 0.4 & 0.45 & 0.48 & 0.5 & 0.8\tabularnewline
\cline{2-8} 
 & \emph{time} & 1.3 & 5.2 & 79 & {*} & 39 & 2.6\tabularnewline
\hline 
\hline 
\multirow{2}{*}{CSMA 4\,4} & $\theta$ & 0.5 & 0.7 & 0.8 & 0.9 & 0.93 & 0.95\tabularnewline
\cline{2-8} 
 & \emph{time} & 0.2 & 0.3 & 4.0 & 8.6 & {*} & 3.8\tabularnewline
\hline 
\hline 
\multirow{2}{*}{WLAN 5} & $\theta$ & 0.1 & 0.15 & 0.18 & 0.2 & 0.25 & 0.5\tabularnewline
\cline{2-8} 
 & \emph{time} & 0.8 & 2.6 & {*} & 2.9 & 2.9 & 1.3\tabularnewline
\hline 
\multirow{1}{*}{WLAN 6} & \emph{time} & 1.3 & 2.2 & {*} & 6.5 & 1.3 & 1.3\tabularnewline
\hline 
\end{tabular}}
\par\end{centering}

\caption{Hypothesis test results for CSMA/CD and WLAN protocols. $\theta$\emph{
}is the threshold probability or the true probability (marked by asterisk).
\emph{time} is simulation time in seconds to achieve the correct result
on a single machine.\label{tab:CSMA}}
\end{table}

\subsection{Choice Coordination}

To demonstrate the scalability of our approach, we consider the choice
coordination model of \cite{NdukwuMcIver2010} and estimate the minimum
probability that a group of six tourists will meet within $T$ steps.
The model has a parameter ($\mathit{BOUND}$) that limits the state
space. We set $\mathit{BOUND}=100$, making the state space of $\approx5\times10^{16}$
states intractable to numerical model checking. Fortunately, it is
possible to infer the correct probabilities from tractable parametrisations.
For $T=20$ and $T=25$ the true minimum probabilities are respectively
$0.5$ and $0.75$. Using smart sampling and a Chernoff bound of $\varepsilon=\delta=0.01$,
we correctly estimate the probabilities to be $0.496$ and $0.745$
in a few tens of seconds on 64 simulation cores.

\subsection{Network Virus Infection\label{sec:virus}}

\begin{figure}
\begin{raggedright}
\begin{minipage}[t]{0.47\columnwidth}%
\begin{center}
\subfloat[Estimates.\label{fig:virusminest}]{\begin{centering}
\includegraphics[angle=-90,width=1\linewidth]{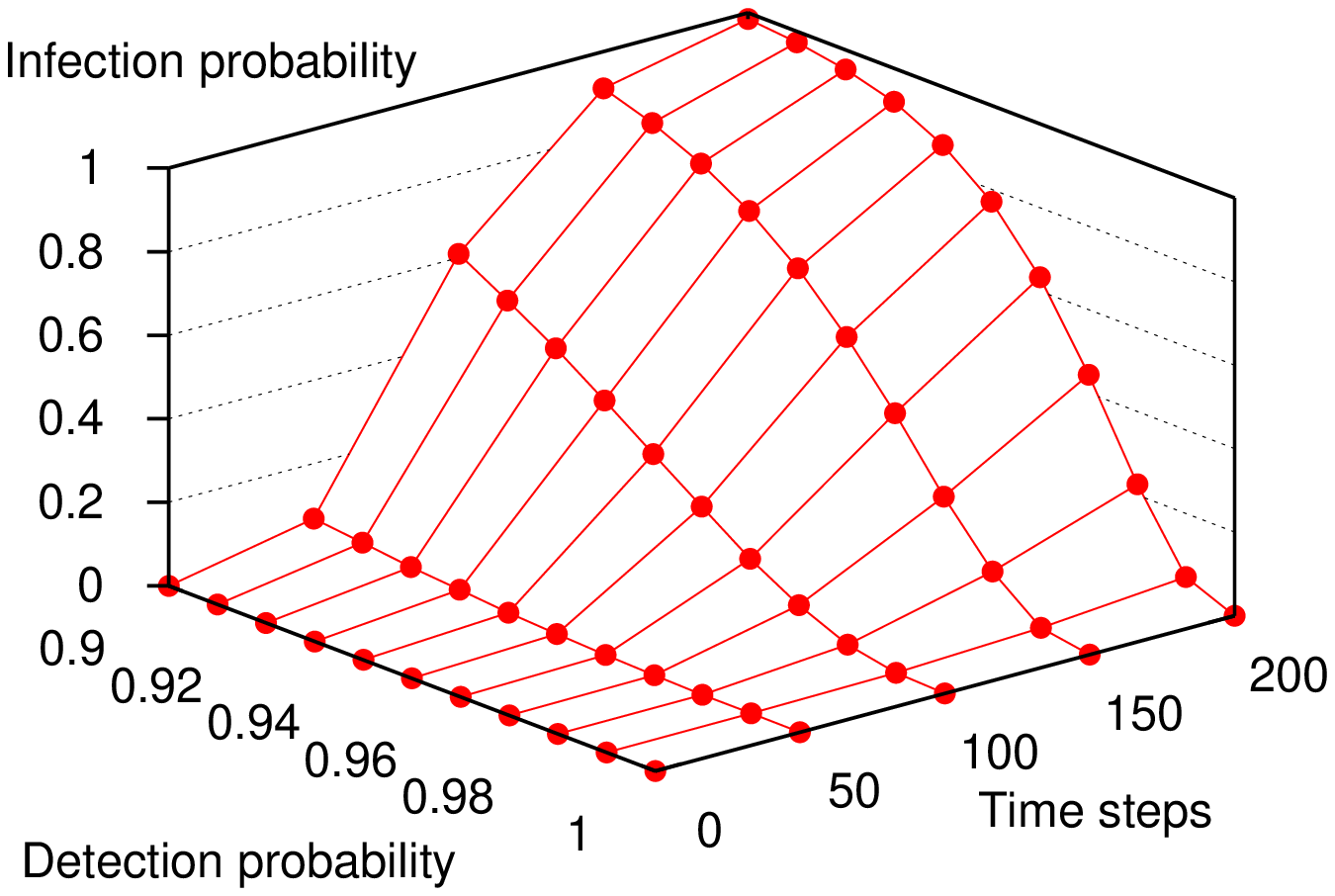}
\par\end{centering}

}
\par\end{center}%
\end{minipage}\quad{}%
\begin{minipage}[t]{0.48\columnwidth}%
\begin{center}
\subfloat[Errors.\label{fig:virusminerr}]{\begin{centering}
\includegraphics[angle=-90,width=1\columnwidth]{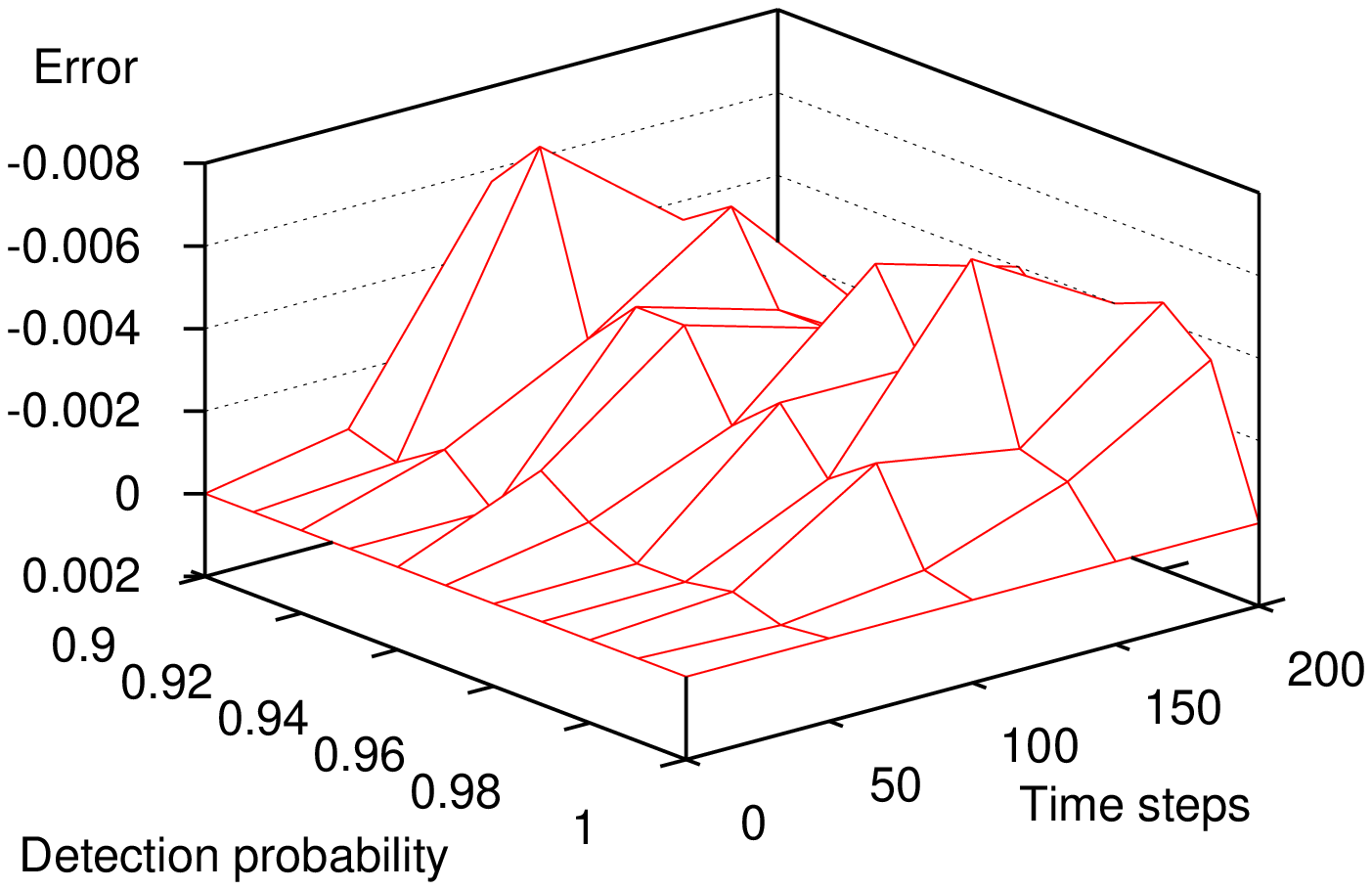}
\par\end{centering}

}
\par\end{center}%
\end{minipage}
\par\end{raggedright}

\caption{Minimum probability of network infection.\label{fig:virusmin}}
\end{figure}

\begin{figure}
\raggedright{}%
\begin{minipage}[t]{0.47\columnwidth}%
\begin{center}
\subfloat[Estimates.\label{fig:virusmaxest}]{\begin{centering}
\includegraphics[angle=-90,width=1\columnwidth]{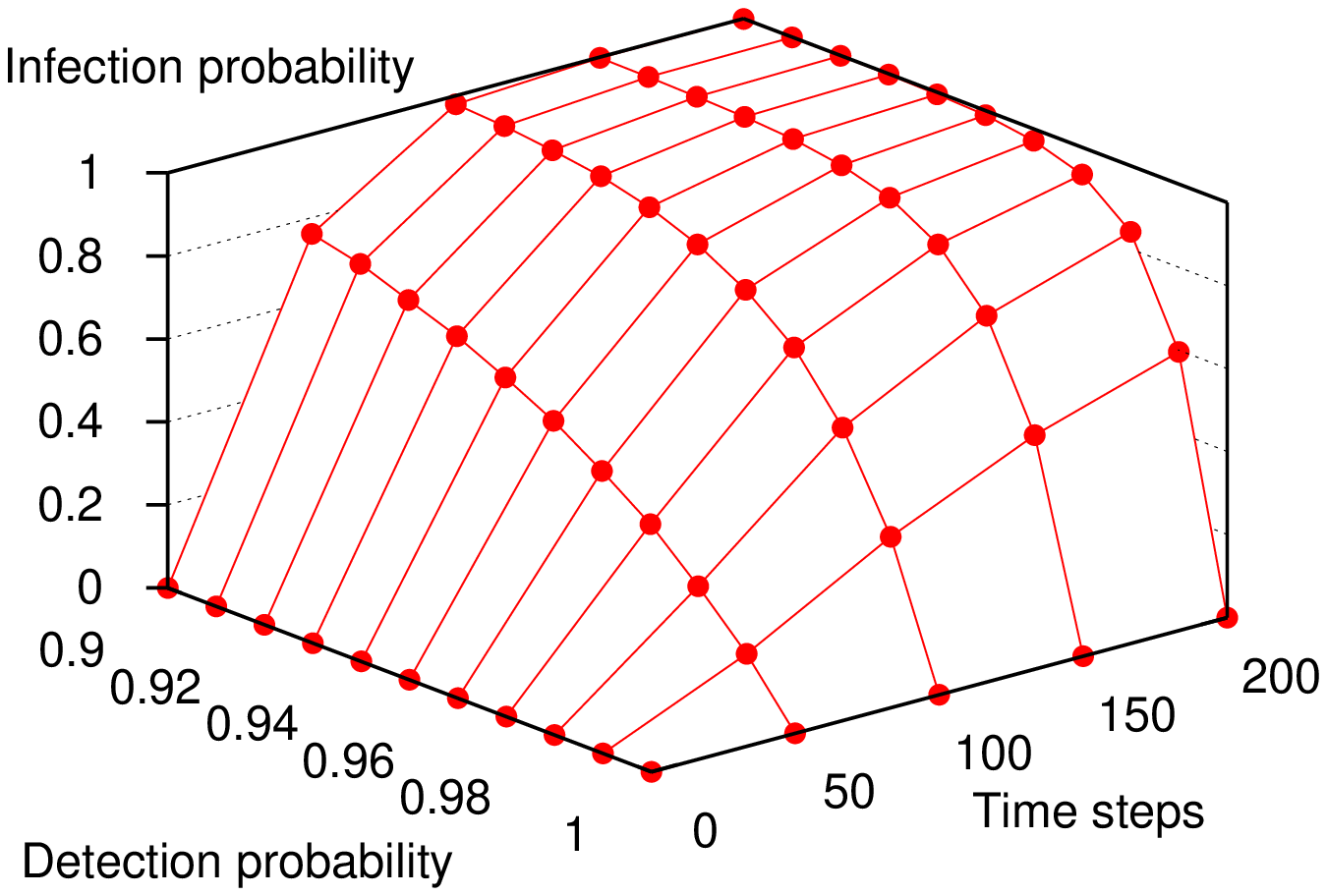}
\par\end{centering}

}
\par\end{center}%
\end{minipage}\quad{}%
\begin{minipage}[t]{0.48\columnwidth}%
\begin{center}
\subfloat[Errors.\label{fig:virusmaxerr}]{\begin{centering}
\includegraphics[angle=-90,width=1\columnwidth]{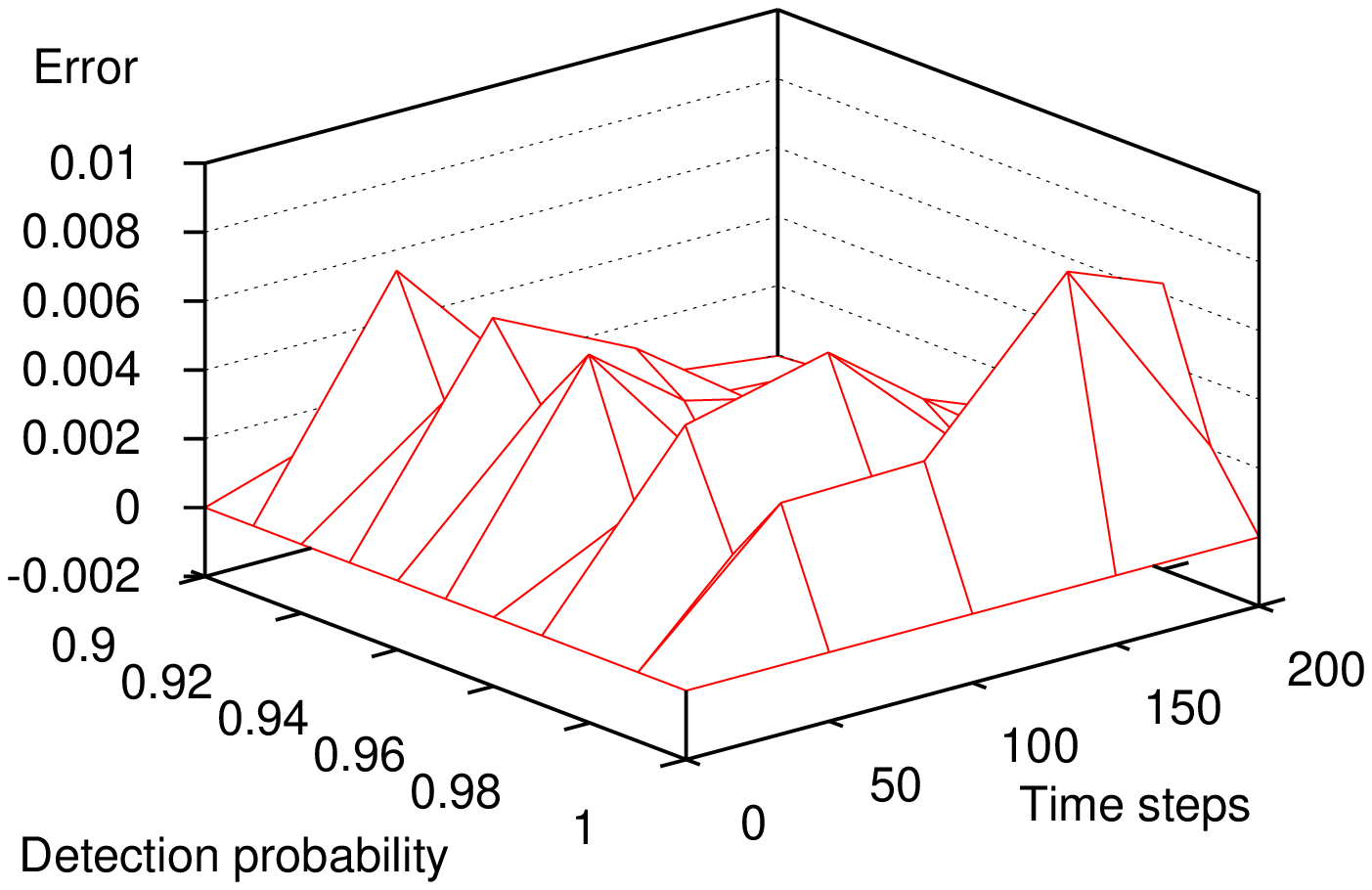}
\par\end{centering}

}
\par\end{center}%
\end{minipage}\caption{Maximum probability of network infection.\label{fig:virusmax}}
\end{figure}

Network virus infection is a subject of increasing relevance. Hence,
using a per-iteration budget of $10^{5}$ simulations, we demonstrate
the performance of Algorithm \ref{alg:smartest} on the \textsc{Prism}
virus infection case study based on \cite{KwiatkowskaNormanParkerVigliotti2009}.
The network is illustrated in Fig. \ref{fig:network} and comprises
three sets of linked nodes: a set of nodes containing one infected
by a virus, a set of nodes with no infected nodes and a set of barrier
nodes which divides the first two sets. A virus chooses which node
to infect nondeterministically. A node detects a virus probabilistically
and we vary this probability as a parameter for barrier nodes. We
consider time as a second parameter. Figs. \ref{fig:virusmin} and
\ref{fig:virusmax} illustrate the estimated probabilities that the
target node in the uninfected set will be infected. We observe in
Figs. \ref{fig:virusminerr} and \ref{fig:virusmaxerr} that the estimated
minimums are within $[-0.0070,+0.00012]$ and the estimated maximums
are within $[-0.00012,+0.0083]$ of their true values. The respective
negative and positive biases to these error ranges reflects the fact
that Algorithm \ref{alg:smartest} converges from respectively below
and above (as illustrated in Fig. \ref{fig:ESTconvergence}). The
average time to generate a point in Fig. \ref{fig:virusmin} was approximately
100 seconds, using 64 simulation cores. Points in Fig. \ref{fig:virusmax}
took on average approximately 70 seconds. 

\subsection{Gossip Protocol\label{sub:gossip}}

\begin{figure}
\begin{centering}
\includegraphics[width=0.55\columnwidth]{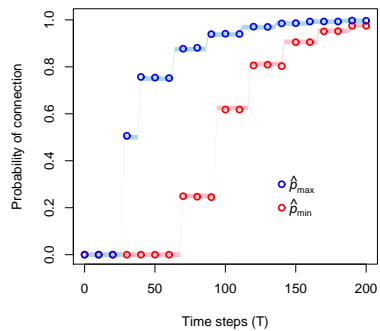}
\par\end{centering}

\caption{Estimated probabilities that maximum path length is $<4$ in gossip
protocol model. Shaded regions denote $\pm0.01$ of true values.\label{fig:gossip}}
\end{figure}

Gossip protocols are an important class of network algorithms that
rely on local connectivity to propagate information globally. Using
the gossip protocol model of \cite{KwiatkowskaNormanParker2008},
we used Algorithm \ref{alg:smartest} with per-simulation budget of
$10^{5}$ simulations to estimate the maximum ($\hat{p}_{\max}$)
and minimum ($\hat{p}_{\min}$) probabilities that the maximum path
length between any two nodes is less than 4 after $T$ time steps.
This is expressed by property $\mathbf{F}^{T}\mathit{max\_path\_len}<4$.
The results are illustrated in Fig. \ref{fig:gossip}. Estimates of
maximum probabilities are within $[-0,+0.0095]$ of the true values.
Estimates of minimum probabilities are within $[-0.007,+0]$ of the
true values. Each point in the figure took on average approximately
60 seconds to generate using 64 simulation cores.

\section{Convergence and Counterexamples\label{sec:counterexamples}}

The techniques described in the preceding sections open up the possibility
of efficient lightweight verification of MDPs, with the consequent
possibility to take full advantage of parallel computational architectures,
such as multi-core processors, clusters, grids, clouds and general
purpose computing on graphics processors (GPGPU). These architectures
may potentially divide the problem by the number of available computational
devices (i.e., linearly), however this must be considered in the context
of scheduler space increasing exponentially with path length. Although
Monte Carlo techniques are essentially impervious to the size of the
state space (they also work with non-denumerable space), it is easy
to construct verification problems for which there is a unique optimal
scheduler. Such examples do not necessarily invalidate the approach,
however, because it may not be necessary to find the possibly unique
optimal scheduler to return a result with a level of statistical confidence.
The nature of the distribution of schedulers nevertheless affects
efficiency, so in this section we explore the convergence properties
of smart sampling and give an example from the literature that does
not converge as well as the case studies in Section \ref{sec:experiments}.

Essentially, the problem is that of exponentially distributed schedulers,
i.e., having a very low mass of schedulers close to the optimum. Fig.
\ref{fig:density} illustrates the difference between exponentially
decreasing and linearly decreasing distributions with the same overall
mass. In both cases $p_{\max}\approx0.2$ (the density at 0.2 is zero),
but the figure shows that there is more probability mass near 0.2
in the case of the linear distribution.

\begin{figure}
\begin{minipage}[t]{0.47\columnwidth}%
\subfloat[Scheduler distributions. Dots denote the results of initial sampling.
The red line is the set of schedulers to refine. Black lines show
the result of subsequent refinements.\label{fig:ECDFconvergence}]{\includegraphics[width=1\textwidth]{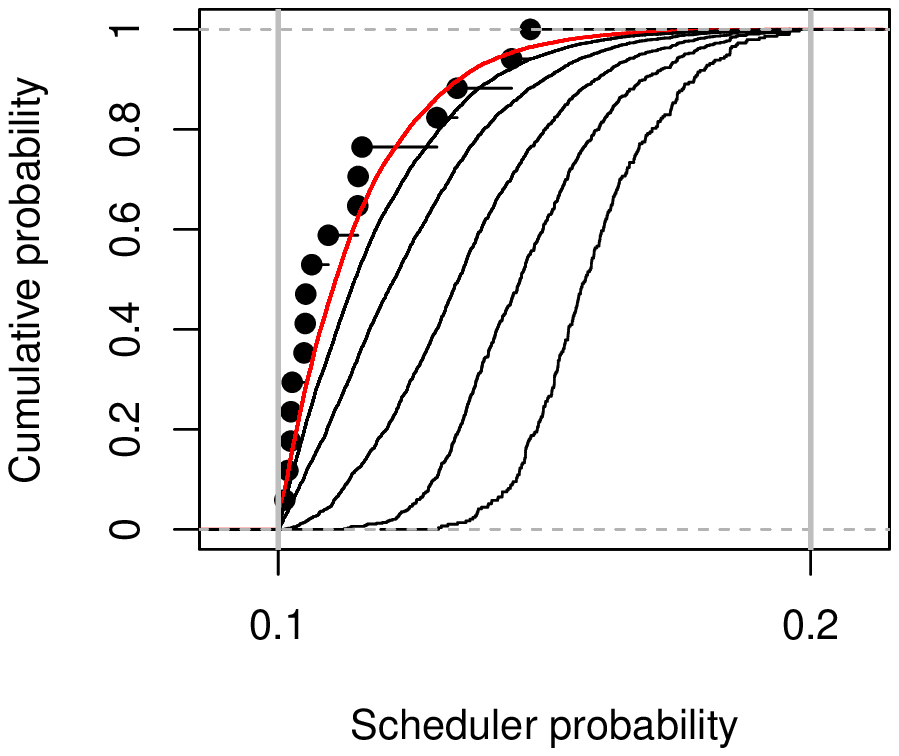}

}%
\end{minipage}\quad{}%
\begin{minipage}[t]{0.47\columnwidth}%
\subfloat[Estimates and schedulers. At each iterative step: $\hat{p}_{\max}$
is the maximum estimate, $\hat{p}_{\mathrm{mean}}$ is the mean estimate
and $\sigma_{\max}$ is the true maximum probability of the available
schedulers.\label{fig:ESTconvergence}]{\includegraphics[width=1\columnwidth]{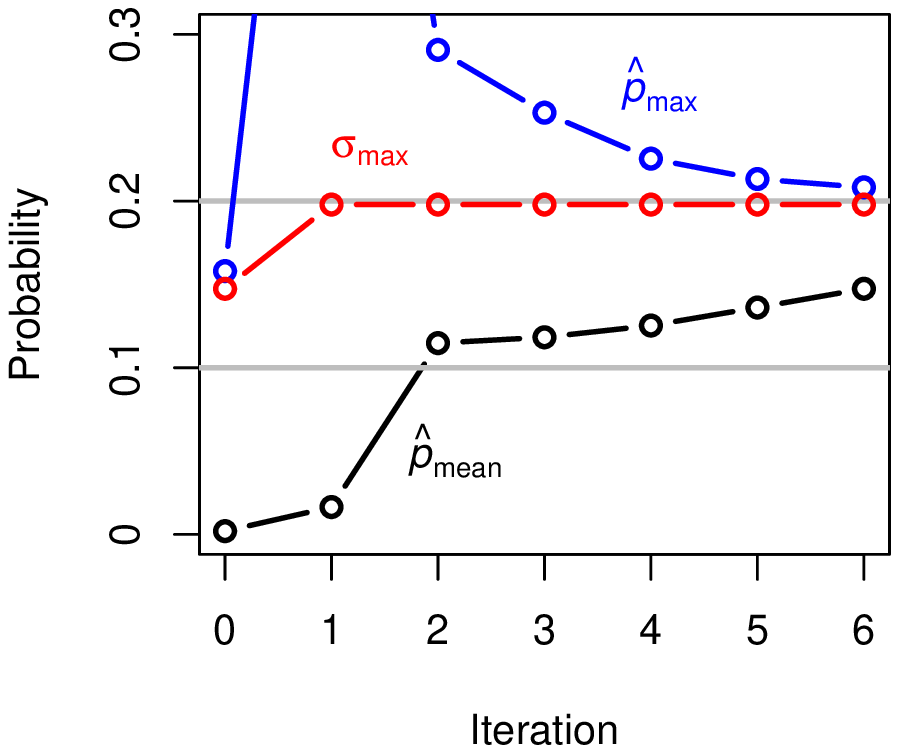}

}%
\end{minipage}

\caption{Convergence of Algorithm \ref{alg:smartest} with exponentially distributed
scheduler probabilities (Fig. \ref{fig:density}) and per-iteration
budget of $10^{6}$ simulations.\label{fig:convergence}}
\end{figure}

\begin{figure}
\begin{minipage}[t]{0.47\columnwidth}%
\begin{center}
\includegraphics[width=1\textwidth]{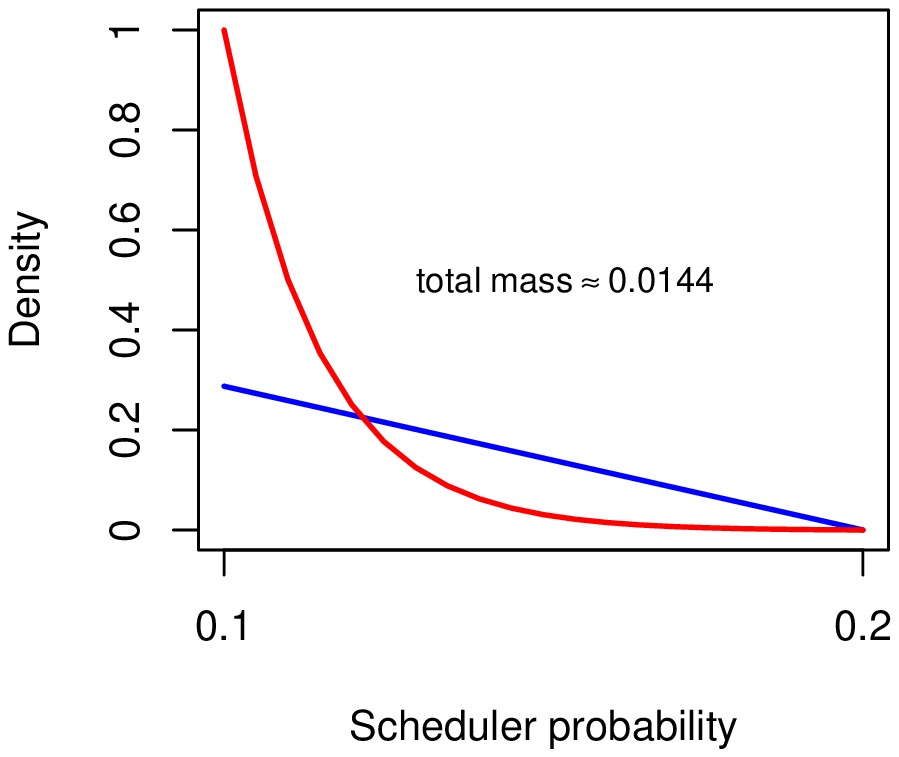}
\par\end{center}

\caption{Theoretical linear (blue) and exponential (red) scheduler densities
with probability mass $\approx0.0144$ and zero density at probability
0.2.\label{fig:density}}
\end{minipage}\quad{}%
\begin{minipage}[t]{0.47\columnwidth}%
\begin{center}
\includegraphics[width=1\columnwidth]{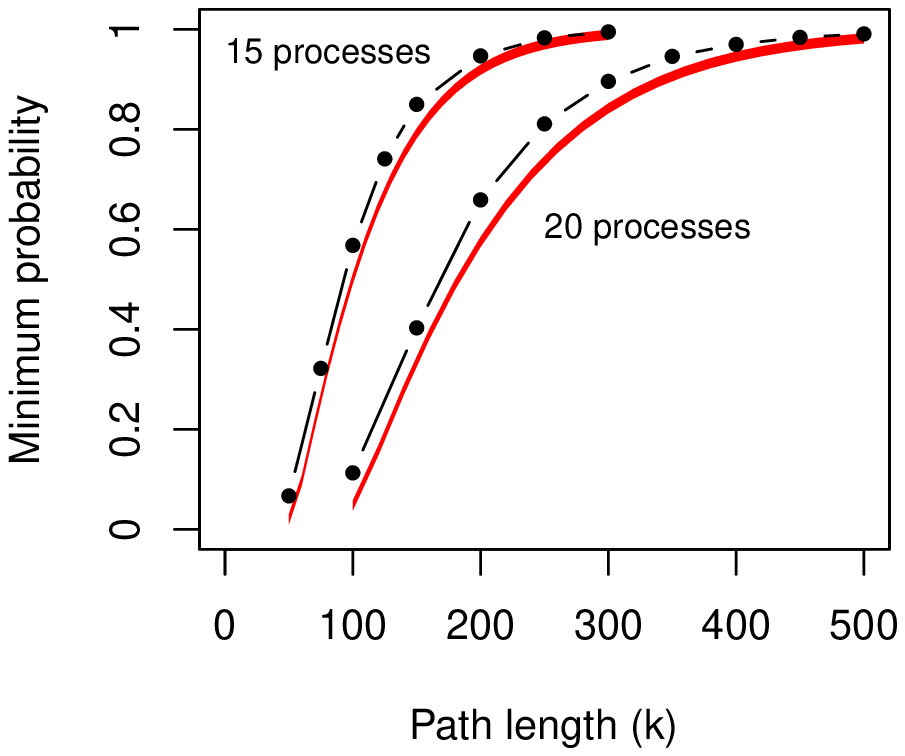}
\par\end{center}

\caption{Performance of smart sampling (black dots) applied to self-stabilising
models of \cite{IsraeliJalfon1990}. Red shaded areas denote true
values $\pm0.01$.\label{fig:self-stab}}
\end{minipage}
\end{figure}

Figure \ref{fig:convergence} illustrates the convergence of Algorithm
\ref{alg:smartest}, using a per-iteration budget of $10^{6}$ applied
to schedulers whose probability of success (i.e., of satisfying a
hypothetical property) is distributed according to the exponential
distribution of Fig. \ref{fig:density}. Fig. \ref{fig:ECDFconvergence}
shows how the initial undirected sampling (black dots) crudely approximates
$p_{\max}$. This approximation is then used to generate the candidate
set of schedulers (red distribution). The black lines illustrate five
iterations of refinement, resulting in a shift of the distribution
towards $p_{\max}$. Fig. \ref{fig:ESTconvergence} illustrates the
same shift in terms of the convergence of probabilities. Iteration
0 corresponds to the undirected sampling. Iteration 1 corresponds
to the generation of the candidate set of schedulers. Note that for
these first two iterations, $\hat{p}_{\mathrm{mean}}$ includes schedulers
that have zero probability of success. The expected value of $\hat{p}_{\mathrm{mean}}$
in these two iterations is equal to the expected probability obtained
by the uniform probabilistic scheduler. This fact can be used to verify
that the hash function and PRNG described in Section \ref{sec:seeds}
sample uniformly. In subsequent iterations the candidates all have
non-zero probability of success. Importantly, the figure demonstrates
that there is a significant increase in the maximum probability of
scheduler success ($\sigma_{\max}$) between iteration 0 and iteration
1, and that this maximum is maintained throughout the subsequent refinements.
Despite the apparently very low density of schedulers near $p_{\max}$,
Algorithm \ref{alg:smartest} is able to make a good approximation.

The theoretical performance demonstrated in Fig. \ref{fig:convergence}
explains why we are able to achieve good results in Section \ref{sec:experiments}.
It is nevertheless possible to find examples for which accurate results
are difficult to achieve. Fig. \ref{fig:self-stab} illustrates the
results of applying Algorithm \ref{alg:smartest} to instances of
the self-stabilising algorithm of \cite{IsraeliJalfon1990}, using
a per-iteration budget of $10^{5}$. Although the estimates (black
dots) do not lie within our statistical confidence bounds of the true
values (red shaded areas), we nevertheless claim that the results
are useful. The problem of quantifying the confidence of estimates
with respect to optimal values remains open, however.

To improve the performance of smart sampling, it is possible to make
an even better allocation of simulation budget. For example, if good
schedulers are very rare it may be beneficial to increase the per-iteration
budget (thus increasing the possibility of seeing a good scheduler
in the initial candidate set) but increase the proportion of schedulers
rejected after each iteration (thus reducing the overall number of
iterations and maintaining a fixed total number of simulations). To
avoid rejecting good schedulers under such a regime, it may be necessary
to reject fewer schedulers in the early iterations when confidence
is low.

\section{Prospects and Challenges \label{sec:prospects}}

The use of sampling facilitates algorithms that scale independently
of the sample space, hence we anticipate that it will be possible
to apply our techniques to nondeterministic models with non-denumerable
schedulers. We believe it is immediately possible to apply smart sampling
to reward-based MDP optimisation problems.

The success of sampling depends on the relative abundance of near
optimal schedulers in scheduler space and our experiments suggest
that these are not rare in standard case studies. While it is possible
to construct pathological examples, where near optimal schedulers
cannot easily be found by sampling, it is perhaps even simpler to
confound numerical techniques with state explosion (three independent
counters ranging over 0 to 1000 is typically sufficient with current
hardware). Hence, as with numerical model checking, our ongoing challenge
is essentially to increase performance and increase the number of
models and problems that may be efficiently addressed. Smart sampling
has made significant improvements over simple sampling, but we recognise
that it will be necessary to develop other techniques to accelerate
convergence. We anticipate that the most fruitful approaches will
be (\emph{i}) to reduce the sampled scheduler space to only those
that satisfy the property and (\emph{ii}) to construct schedulers
piecewise. Such techniques will also reduce the potential of hash
function collisions.

An important remaining challenge is to quantify the confidence of
our estimates and hypothesis tests with respect to optimality. In
the case of hypothesis tests that satisfy the hypothesis, the statistical
confidence of the result is sufficient. If an hypothesis is not satisfied,
however, the statistical confidence does not relate to whether there
exists a scheduler to satisfy it. Likewise, the statistical confidence
bounds of probability estimates imply nothing about how close they
are to the true optima. We nevertheless know that our estimates of
the extrema must lie within the true extrema or exceed them with the
specified statistical confidence. This is already useful and a significant
improvement over the results produced using the uniform probabilistic
scheduler. In addition, given the number of simulations performed,
we may at least quantify confidence with respect to the product $p_{g}p_{\overline{g}}$
(the rarity of near optimal schedulers times the average probability
of the property with near optimal schedulers).

\section*{Acknowledgements}

We are grateful to Beno\^it Delahaye for useful prior discussions.
This work was partially supported by the European Union Seventh Framework
Programme under grant agreement no. 295261 (MEALS).

\bibliographystyle{plain}
\bibliography{SmartSampling}

\end{document}